\documentclass[11pt,a4paper]{article}
\pdfoutput=1
\usepackage{jheppub}

\usepackage{graphicx}
\usepackage{epsfig}
\usepackage{bm}
\usepackage{latexsym,amssymb,amsmath,amssymb,wasysym,float,comment}

\usepackage{color}

\usepackage{scrextend}
\usepackage{mathrsfs}
\usepackage{enumitem}

\usepackage[usenames,dvipsnames]{xcolor}
\definecolor{orange}{cmyk}{0,0.5,1,0}
\definecolor{rossoCP3}{cmyk}{0,.88,.77,.40}
\definecolor{graa}{rgb}{0.8,0.8,0.8}
\definecolor{blaa}{rgb}{0.2,0.2,0.6}

\begin{document}

\begin{flushright}
MPP-2025-113\\
LMU-ASC 15/25
\end{flushright}

\title{Bulk/boundary Modular Quintessence and DESI}

\author[a,b,c]{Luis A. Anchordoqui,}

\affiliation[a]{Department of Physics and Astronomy,  Lehman College, City University of
  New York, NY 10468, USA
}

\affiliation[b]{Department of Physics,
 Graduate Center,  City University of
  New York,  NY 10016, USA
}

\affiliation[c]{Department of Astrophysics,
 American Museum of Natural History, NY
 10024, USA
}

\author[d,e]{Ignatios Antoniadis,}

\affiliation[d]{High Energy Physics Research Unit, Faculty of Science, Chulalongkorn University, Bangkok 1030, Thailand}

\affiliation[e]{Laboratoire de Physique Th\'eorique et Hautes \'Energies
  - LPTHE \\
Sorbonne Universit\'e, CNRS, 4 Place Jussieu, 75005 Paris, France
}

\author[f]{Niccol\`o Cribiori,}

\affiliation[f]{KU Leuven, Institute for Theoretical Physics, Celestijnenlaan 200D, B-3001 Leuven, Belgium}

\author[g]{Arda Hasar,}

\affiliation[g]{Department of Physics, Faculty of Arts and Sciences,
Middle East Technical University, 06800 Ankara, T\"urkiye}

\author[h,i]{Dieter~L\"ust,}

\affiliation[h]{Max--Planck--Institut f\"ur Physik,  
 Werner--Heisenberg--Institut,
Boltzmannstra{\ss}e 8, 85748 Garching, Germany
}

\affiliation[i]{Arnold Sommerfeld Center for Theoretical Physics, 
Ludwig-Maximilians-Universit\"at M\"unchen,
80333 M\"unchen, Germany
}

\author[h]{Joaquin Masias,}

\author[j,k]{Marco Scalisi}

\affiliation[j]{Department of Physics and Astronomy ``Ettore Majorana'', University of Catania, Via S. Sofia 64, 1-95125 Catania, Italy}

\affiliation[k]{INFN-Sezione di Catania, Via Santa Sofia 64, I-95123 Catania, Italy}

\abstract{ The latest DESI DR2 results, when combined with other independent cosmological data on the Cosmic Microwave Background and supernovas, suggest a preference for dynamical dark energy. 
 We propose a novel cosmological scenario, which features two distinct scalar fields. One governs the magnitude of the present-day dark energy density and is related to the size of extra-dimensions. Accounting for the observed smallness of this energy density requires the scalar to reside near the boundary of field space. 
 The second field,  responsible for the time evolution of dark energy and associated with the string coupling, must instead lie in the bulk to remain consistent with the non-observation of light string states. We show that a natural candidate for such dark energy dynamics is a quintessence modular-invariant potential, in which the second scalar field rolls down a negatively curved slope, starting from a self-dual critical point. We find that this scenario is in good agreement with the latest findings by DESI.

}

\maketitle

\section{Introduction}

The latest findings on baryon acoustic oscillations (BAO) from the Dark Energy Spectroscopic Instrument (DESI) data release 2 (DR2)~\cite{DESI:2025zgx} combined with CMB information and supernova (SN) datasets (PantheonPlus~\cite{Brout:2022vxf}, Union3~\cite{Rubin:2023ovl}, and DESY5~\cite{DES:2024jxu}) suggest a preference for dynamical dark energy over a cosmological constant (see also \cite{Sabogal:2025jbo}). Although this is not statistically significant yet, it is interesting to entertain the possibility that the DESI DR2 results correspond to a real signal of physics beyond $\Lambda$CDM.

From the theory side, de Sitter vacua, as a way to realize a pure cosmological constant, have consistently proven challenging to reconcile with generic properties of quantum gravity \cite{Maldacena:2000mw,Kachru:2003aw,Dvali:2014gua,Dvali:2017eba,Obied:2018sgi,Dvali:2018fqu,Garg:2018reu,Ooguri:2018wrx,Bena:2018fqc,Dvali:2018jhn,Bena:2020xrh,Lust:2022lfc}, 
prompting the search for alternative  scenarios. Concretely, a simple mechanism for dynamical dark energy involves a non-interacting quintessence field slowly rolling down its potential~\cite{Peebles:1987ek,Ratra:1987rm,Wetterich:1994bg,Caldwell:1997ii}. An option, analyzed by the DESI Collaboration~\cite{DESI:2025fii}, is given by an axion-like potential with a scalar field in its `hilltop regime' slowly rolling down. On the other hand, fast rolling potentials, such as steep exponential functions, are typically disfavored for quintessence \cite{Bhattacharya:2024hep, Bhattacharya:2024kxp}; see however the recent discussion in~\cite{Andriot:2025los}.

Self-dual potentials have recently emerged as an interesting description of thawing quintessence, with results that align well with the DESI DR2 data, as it was shown in~\cite{Anchordoqui:2025fgz}. Such scenarios are motivated by the existence of duality symmetries in string theory, most notably $S$-duality, as first proposed in~\cite{Font:1990gx}.
In the present paper, we will extend the discussion of~\cite{Anchordoqui:2025fgz} to the case of potentials invariant under the full modular group SL$(2,{\mathbb Z})$, which arises naturally in string compactifications as target space modular invariance. 
The implications of imposing modular invariance on the resulting low-energy effective actions have been extensively studied in  early foundational works~\cite{Ferrara:1989bc,Font:1990nt,Font:1990gx,Cvetic:1991qm}, reviewed in \cite{Cribiori:2024qsv}. More recently, modular invariant inflationary and de Sitter models have been studied e.g.~in \cite{Leedom:2022zdm,Casas:2024jbw,Kallosh:2024ymt,Kallosh:2024pat}.
Concretely, we will consider the {modular-invariant potential}
\begin{equation}
\label{Vintro}
V (S, \bar S)  \simeq -\frac{1}{\log [|\eta(S)|^{4} \ (S+\bar  S)]}    \,, 
\end{equation}
in terms of the complex field $S$. As we will show in Sec.~\ref{Sec:ModPotential}, the above potential, when truncated along its saxionic direction, shares several features of the self-dual potential of~\cite{Anchordoqui:2025fgz}. In particular, it satisfies swampland constraints, which are quantum gravity consistency conditions, and it provides a very good fit to the DESI DR2 data.
 
An account for the time evolution of the dark energy density, however, does not provide an explanation for the remarkable smallness of its observed value today. This fundamental problem remains one of the most significant challenges in theoretical physics and continues to motivate ongoing research.
Assuming the anti-de Sitter distance conjecture~\cite{Lust:2019zwm} extends to positive energy backgrounds, the smallness of the cosmological constant would imply the existence of a light tower of states, $e.g$.~associated to light Kaluza Klein modes that belong to a large internal compact space. This is the reasoning behind the Dark Dimension scenario \cite{Montero:2022prj}, in which the dark energy density is related to the radius $R$ of a single, micron-sized extra dimension by $\Lambda(R)\simeq R^{-4}$.

Motivated by these considerations, in Sec.~\ref{Sec:HybScenario} we propose a {novel hybrid scenario} involving two scalar fields. The first is the size $R$ of the extra dimension, which lies near the weakly coupled, asymptotic boundary of moduli space. The second is $s \equiv (S + \bar{S})/2$, which drives the quintessence dynamics and, therefore, accounts for the time variation of dark energy, as suggested by the DESI DR2 data. 
We identify $s$ with the dilaton, the scalar field that determines the string coupling, but more generally it could represent another geometric modulus.
In our scenario, quintessence takes place in the vicinity of the self-dual point $s=1$, and therefore in the bulk of the moduli space. Conversely, the smallness of the present-day dark energy density arises from $R$ residing in an asymptotic region. This {bulk/boundary} scenario can thus be realized by a potential that factorizes into an $R$- and an $s$-dependent term, such as 
\begin{equation}
\label{VRsintro}
V(R,s) = \Lambda(R) \, V_q(s)\,,
\end{equation}
where $\Lambda(R)\simeq R^{-4}$ and $V_q(s)$ can be given by \eqref{Vintro} for $S=\bar{S}$.

In Sec.~\ref{Sec:DESI}, we confront the predictions of our scenario with the latest DESI DR2 data and find them in good agreement.  In Sec.~\ref{Sec:SUGRA}, we briefly explore possible supersymmetrizations of Eq.\eqref{VRsintro}, making use of the versatility offered by non-linearly realized supersymmetry.  We also outline the main challenges involved in achieving a full-fledged embedding of this framework into more fundamental theories, such as supergravity and string theory. We leave this ambitious task for future work.

\section{A modular-invariant potential}\label{Sec:ModPotential}

\begin{figure}[t]
    \centering
        \begin{tabular}{cc}
        \includegraphics[width=0.48\linewidth]{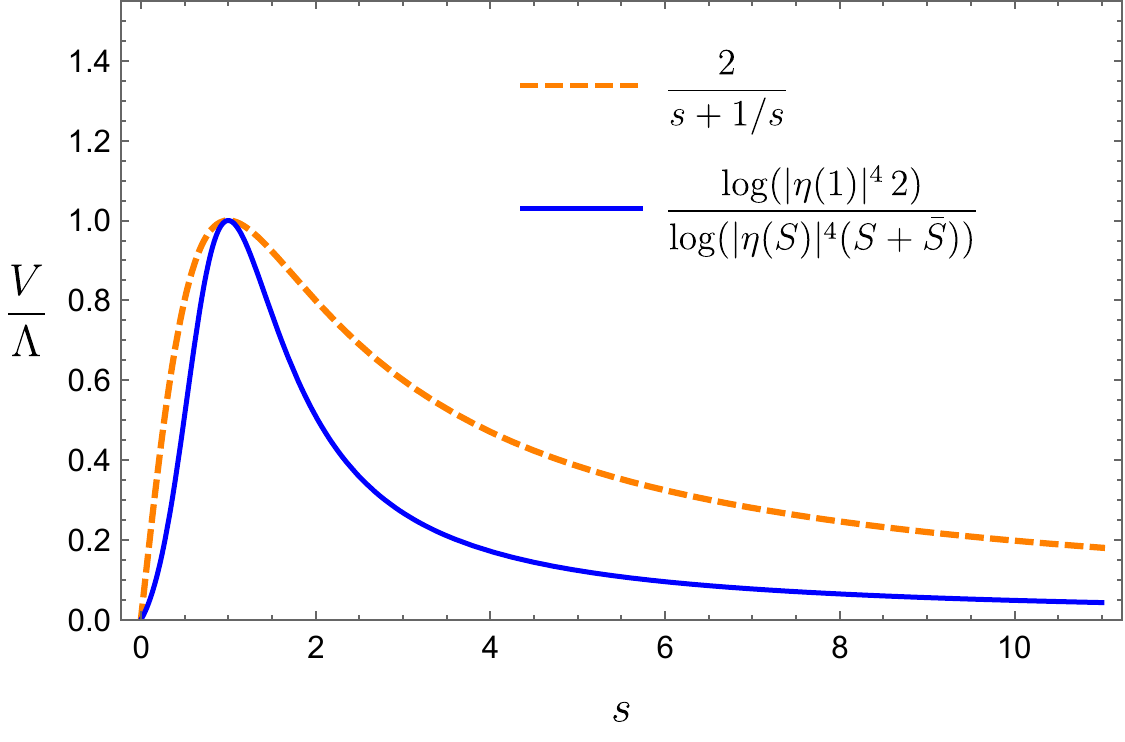}   & \includegraphics[width=0.48\linewidth]{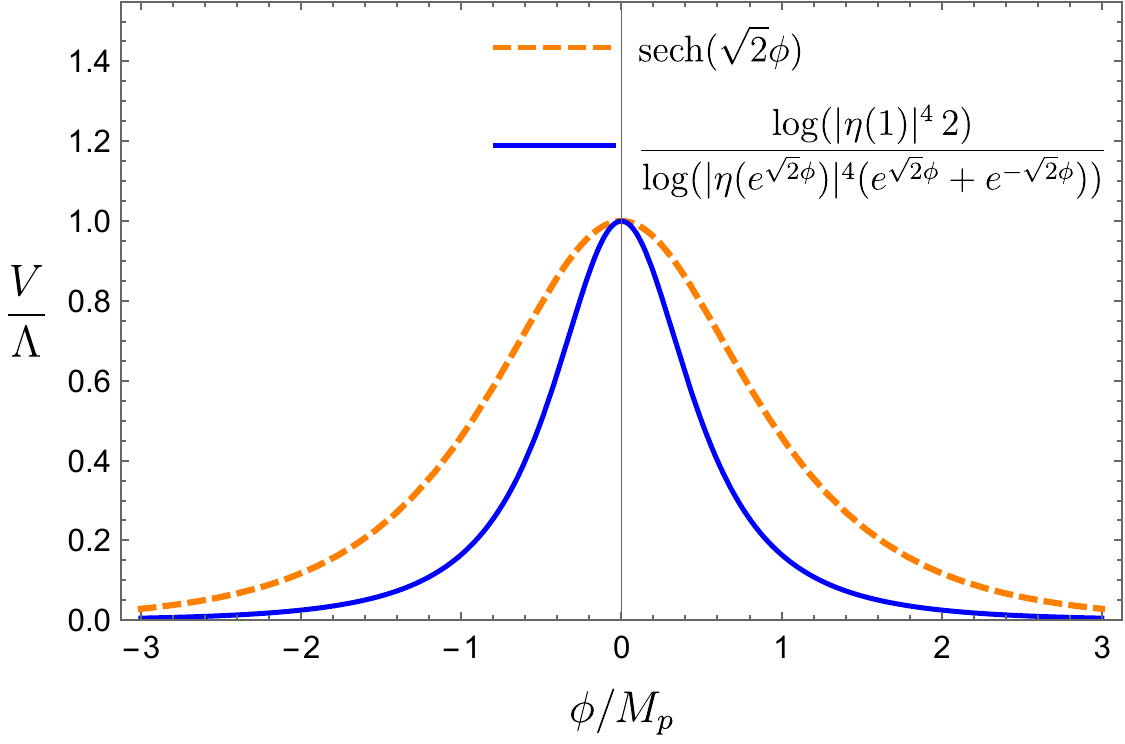}
        \end{tabular}
   \vspace{-0.2cm} \caption{Modular-invariant potential of eq.~(\ref{Vguess})
      (blue) compared to the S-dual potential of eq.~(\ref{Vinit})
      (orange) in the $s$ (left) and $\phi$ (right) coordinates.}
    \label{fig:Vcomparison}
\end{figure}

In this section, we present the modular-invariant potential around which this article is centered.
Our starting point is the S-dual scalar potential proposed in \cite{Anchordoqui:2025fgz}, namely  
\begin{equation}
\label{Vinit}
V = \Lambda \, \text{sech}\left(\sqrt 2\phi/M_p\right) =    \frac{{2 \ \Lambda}}{s+1/s},
\end{equation}
with $s=e^{\sqrt2 \phi/M_p}$ a real scalar field and $\Lambda$ a free parameter. In this model, $\phi$ plays the role of the quintessence field and is assumed to be canonically normalized. We set the free parameter equal to the value of the dark energy density measured today, namely $\Lambda \sim 10^{-120}M_p^4$; we will discuss in the next section how $\Lambda$ should also be regarded as a field-dependent quantity. The plot of the potential \eqref{Vinit} can be seen as a dotted orange line in figure \ref{fig:Vcomparison}.

As argued in~\cite{Anchordoqui:2025fgz}, the potential (\ref{Vinit}) has a number of attractive features, which we summarize below. First, it is compatible with a number of quantum gravity constraints, such as swampland conjectures. Indeed, around the $\mathbb{Z}_2$-symmetric (self-dual point) point $\phi=0$ ($s=1$), we have
\begin{equation}
    \frac{V'}{V}\bigg\vert_{\phi=0}=0 \qquad {\rm and} \qquad \frac{V''}{V}\bigg\vert_{\phi=0}=-\frac{2}{M_p^2} ,
\end{equation}
where primes denote derivatives with respect to $\phi$. These relations are in agreement with the (refined) de Sitter conjecture \cite{Obied:2018sgi,Garg:2018reu,Ooguri:2018wrx}. Instead, asymptotically we have
\begin{equation}
\frac{V'}{V}\bigg\vert_{\phi\rightarrow\infty}=-\frac{\sqrt{2}}{M_p},
\end{equation}
saturating the TCC bound \cite{Bedroya:2019snp}. 
Second, the potential can accommodate the new DESI results~\cite{DESI:2025fii} near the $\mathbb{Z}_2$-symmetric point. More precisely, the potential changes concavity at $\phi^*/M_P = \frac{\sqrt 2}{2}\log(\pm 1 + \sqrt{2}) \simeq 0.62$. This is consistent with  DESI DR2 results~\cite{DESI:2025fii} for which quintessence occurs around $\phi/M_p \sim 0.5$.
Third, the potential describes the cooling off of the universe as a consequence of spontaneous breaking of the $\mathbb{Z}_2$ symmetry. Hence, quintessence occurs via a scalar starting at the top of the hill, by virtue of the symmetry enhancement \cite{Chen:2025rkb}, and then rolling down the slope.

\begin{figure}[t]
    \centering
        \begin{tabular}{cc}
        \includegraphics[width=0.5\linewidth]{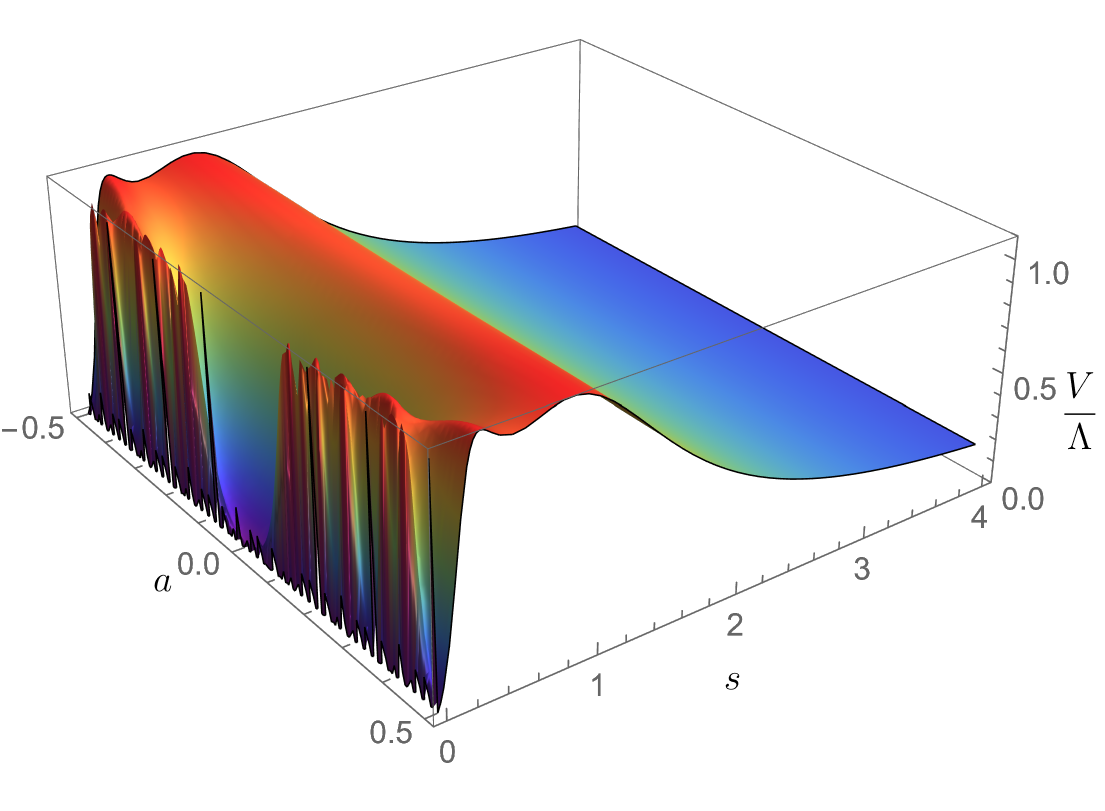}   & \includegraphics[width=0.46\linewidth]{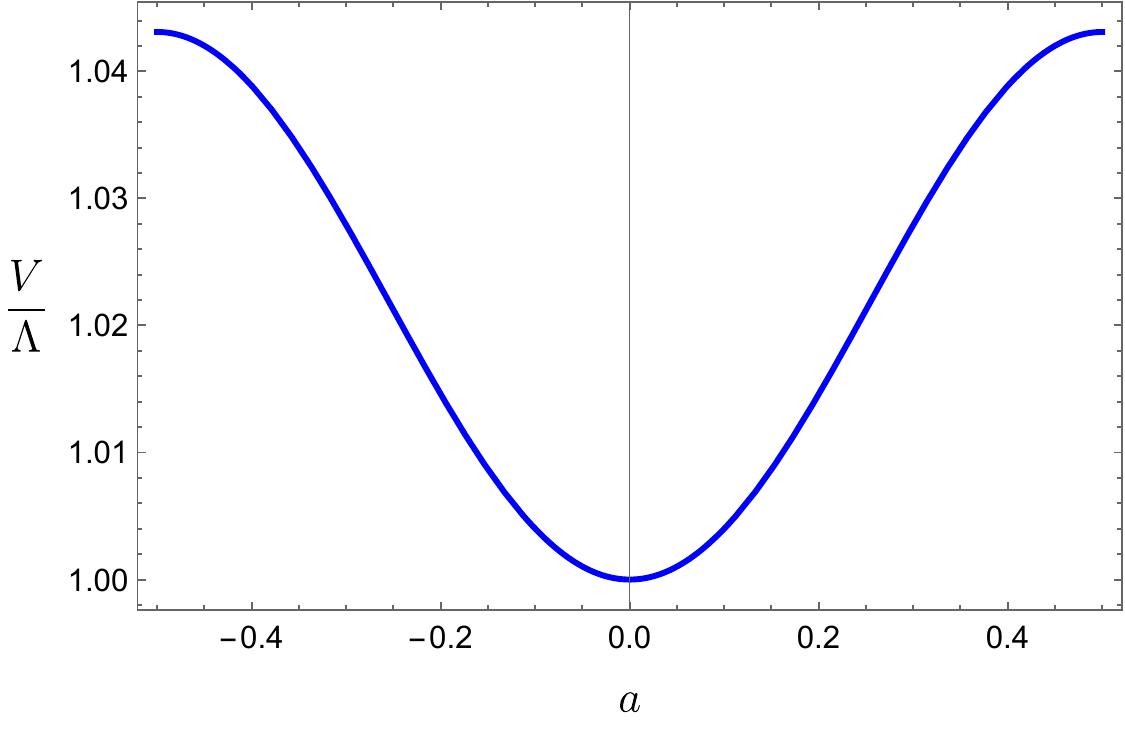}
        \end{tabular}
   \vspace{-0.2cm} \caption{Left panel: plot of the modular invariant potential in the $s$-$a$ plane. Right panel: potential along the axionic direction for $s=1$. At $a=0$, $s=1$ the potential has a minimum which persists for $s \geq 1$.}
      \label{fig:V2d}
\end{figure}

The main property of the scalar potential \eqref{Vinit} is that it is invariant under $\phi \to -\phi$, or more suggestively $s\to 1/s$. We want to extend this to an invariance under the full modular group. For this purpose, we first promote $S$ to a complex scalar field
\begin{equation}
S = s+ia= e^{\tilde \phi}+ia\,, \qquad \text{with} \qquad \tilde \phi=\sqrt 2 \phi/M_p\,,
\end{equation}
transforming under SL$(2,\mathbb{Z})$ as
\begin{equation}
\label{eq:modtranf}
S \to \frac{aS +ib}{icS+d}\,.
\end{equation} 
Then, in order to construct a modular invariant-function of this
complex field, we recall that the combination
\begin{equation}
\label{eq:modinfcomb}
|\eta(S)|^{4} \ (S+\bar S)\,,
\end{equation}
where $\eta(S)$ is the Dedekind $\eta$-function, is modular-invariant. It exhibits the asymptotics
\begin{align}
{\rm Re} \ S \to \infty \qquad &\eta(S) \to e^{-\frac{\pi}{12}{\rm Re}S}\,,\\
{\rm Re} \ S \to \infty \qquad &|\eta(S)|^{4} \ (S+\bar S) \to
e^{-\frac{n\pi}{3}{\rm Re} \, S}(S+\bar S)\,.
\end{align}
Then, an educated guess for the modular-invariant scalar potential is
\begin{equation}
\label{Vguess}
V(S, \bar S)  = -\frac{ \Lambda \ \log\left[|\eta(1)|^{4} \, 2\right]}{\log [|\eta(S)|^{4} \ (S+\bar  S)]} \, ,
\end{equation}
where the overall factor is chosen so to give  $V(1, 1)= \Lambda$ at the self-dual point (recall that $\eta(1)=\Gamma \left(1/4\right)/(2 \pi ^{3/4})$). This potential exhibits several features in common with the S-dual function \eqref{Vinit} discussed above, including similar behavior near the self-dual point ($s=1$) and in the asymptotic regime ($s \to \infty$). Its profile in the $a$–$s$ plane is illustrated in figure~\ref{fig:V2d}. Modular invariance is especially evident in the region $s < 1$, where infinitely many copies of the fundamental domain appear. Along the axionic direction $a$, the potential exhibits an oscillatory behavior. Notice however that the effect is relatively small, about 4\%, which makes the oscillations at $s = 1$ difficult to discern in the left panel of figure~\ref{fig:V2d}. Notably, $a = 0$ constitutes a consistent truncation, as the potential has a minimum there. The right panel of figure~\ref{fig:V2d} clearly shows this minimum along the $a$-direction at the self-dual point, which in fact persists for all values of $s \geq 1$.

The potential \eqref{Vguess} is the main formula around which this work is centered. It is a special case of a more general class recently studied in \cite{Kallosh:2024ymt} to build suitable inflation potentials with predictions compatible with the latest CMB data. Our motivation is however different from \cite{Kallosh:2024ymt}, since we are interested in late-time cosmology. It would be interesting to study if also other models in the same class can be employed for quintessence. We leave this investigation for future work.

In the following, we will explain how this modular-invariant potential can be motivated from general principles of quantum gravity. Then, we will compare the potential in detail with the DESI DR2 data.

\section{A novel scenario: dark energy from bulk and asymptotics }\label{Sec:HybScenario}

In this section,  we provide the theoretical background and motivation behind the modular-invariant potential introduced above. We first review general facts about the ultraviolet cutoff and towers of states in quantum gravity. These motivate our proposal of a novel scenario where the time-evolution of dark energy is due to the presence of a tower of string modes in the bulk of moduli space, and therefore related to the strength of the string coupling $g_s$, while its (extremely small) magnitude is related to the presence of one or more mesoscopic extra dimensions (see e.g. \cite{Montero:2022prj,Anchordoqui:2023oqm,Anchordoqui:2025nmb}). We then explain how this scenario can be linked to the modular-invariant potential of the previous section.

\subsection{Species scale and towers}

An upper bound on the ultraviolet cutoff is given by the species scale $\Lambda_{sp}$  \cite{Veneziano:2001ah, Dvali:2007hz,Dvali:2009ks}.
Its computation is in general non-trivial, but it can at present be performed in some simplified setups and under specific assumptions. One approach is to compute the number $N_{sp}$ of states with mass below $\Lambda_{sp}$, giving then
\begin{equation}
\Lambda_{sp}=\frac{M_p}{{(N_{sp})^{\frac{1}{d-2}}}}\ .
\label{speciesscale}
\end{equation}
This computation is challenging \cite{Castellano:2022bvr,Blumenhagen:2023yws}, and it is unclear how to perform it away from asymptotic regions of the moduli space, where the concept of mass is ill-defined \cite{Long:2021jlv}.
Another possibility is to read-off $\Lambda_{sp}$ from the scale controlling higher-curvature corrections to the (gravitational sector of) the effective action. For example, in the effective action of type II string theory compactified on Calabi-Yau down to four dimensions, the quadratic curvature terms are controlled by the one-loop topological free energy $F_1$ \cite{Bershadsky:1993ta,Antoniadis:1993ze}, and the species scale can be expressed as \cite{vandeHeisteeg:2022btw,Cribiori:2022nke,vandeHeisteeg:2023ubh}
\begin{equation}
\Lambda_{sp}={M_p\over\sqrt{F_1}}\, \label{f1}\, .
\end{equation}
Matching between the two definitions above requires that the index-like quantity $F_1$ is a proxy for the number of species, $N_{sp} \simeq F_1$. 
Yet another possibility is to define the species scale as the inverse radius of the smallest possible black hole in the effective description \cite{Dvali:2007hz,Dvali:2007wp,Cribiori:2022nke, Calderon-Infante:2025pls}, leading to 
\begin{equation}
\Lambda_{sp}=\frac{M_p}{{({\cal S}_{\rm min})^{\frac{1}{d-2}}}}\, ,
\end{equation}
where $\mathcal{S}_{min}$ is the entropy of such a black hole. This definition identifies the number of species $N_{sp}$ with the intensive black hole entropy, $N_{sp} \simeq \mathcal{S}_{min}$. It has subsequently led to the notion of species thermodynamics \cite{Cribiori:2023ffn,Basile:2024dqq,Herraez:2024kux}.

While not always completely equivalent, all of these definition prelude at the fact that $\Lambda_{sp}$ should be a function of scalar fields in such a way that, in asymptotic infinite distance limits on the moduli space, where an infinite tower of states becomes light, also the species scale vanishes parametrically. 
According to the emergent string conjecture \cite{Lee:2019wij}, these light states emerging at the boundary of the moduli space can be either Kaluza-Klein or string excitations. In a decompactification limit of $p$ dimensions of size $R$, one has asymptotically 
\begin{eqnarray}
\Lambda_{sp}\simeq R^{-\frac{p}{d+p-2}} M_p^{\frac{d-2}{d+p-2}} \simeq \frac{M_p}{ (M_pR)^{\frac{p}{d+p-2}}}\, ,  \qquad i.e. \qquad N_{sp}\simeq (M_pR)^{\frac{(d-2)p}{d+ p-2}}\, ,
\end{eqnarray}
which corresponds to the $(d+p)$-dimensional Planck mass, $\hat M_{p}$. 
In an emergent string limit, one has asymptotically
\begin{equation}
\Lambda_{sp} \simeq \hat g_s^{\frac{2}{d+p-2}} \hat M_p\, ,  \qquad i.e. \qquad N_{sp} \simeq \hat g_s^{-\frac{2(d-2)}{d+p-2}} (M_p R)^{\frac{p(d-2)}{d+p-2}} = g_s^{-2},
\end{equation}
where $\hat g_s$ is the $(d+p)$-dimensional string coupling, while $g_s$ the $d$-dimensional one. Hence, in this limit one has $\Lambda_{sp} \simeq  M_s = g_s^{\frac{2}{d-2}} M_p$ . Note that the dilaton $s=g_s^{-2}$ will be later identified with the saxionic field of the modular invariant model presented above. 

Those just discussed are the prototype cases of a pure limit along which either a Kaluza-Klein or a string tower dominates. The species scale can then be expressed in terms of only one of the two towers. In a more general situation, one can simultaneously move along the infinite distance limit in both the $g_s$ and $R$ directions (see $e.g.$~\cite{Etheredge:2024tok,Etheredge:2025ahf} for a recent discussion on infinite distance limits and their classification), while still being compatible with the emergent string conjecture. 
For our purposes, we parametrize this scenario asymptotically as
\begin{eqnarray}
\Lambda_{sp}\simeq g_s^q \frac{M_p}{(M_pR)^{\frac{p}{d+p-2}}}\, , \qquad{ i.e. }\qquad N_{sp}\simeq g_s^{-q(d-2)}(M_pR)^{\frac{p(d-2)}{d+p-2}}\, ,\label{speciesscalefinal}
\end{eqnarray}
where we introduced a mixing parameter $0 \leq q \leq 2/(d-2)$. We can further bound these functions by asking that $1/R \leq M_s\leq\Lambda_{sp}$, leading to\footnote{We thank I.~Basile for discussion on this point.}
\begin{equation}
\label{relationgs}
\frac{1}{M_p R} \leq g_s \leq \left(\frac{1}{M_p  R}\right)^{\frac{p}{(d+p-2)(1-q)}}.
\end{equation}
Notice that the self-dual point, $g_s=1$, is not contained in this region, since $M_s < M_p$.

\subsection{Quintessence and dark dimensions}

Motivated by the discussion presented above, we now consider a scenario in which the scalar potential is a function of $g_s$ and $R$ and discuss how this can arise below.

In any consistent gravitational effective theory, a general requirement is that the scale set by the potential has to be smaller than the ultraviolet cutoff. This is especially the case when one can use $\sqrt{V/M_p^2}$ as a proxy for the infrared cutoff, as for example in a positive energy phase with a cosmic horizon. The requirement that $V \leq M_p^2 \Lambda_{sp}^2$ \cite{Hebecker:2018vxz,Scalisi:2018eaz} is in fact necessary to ensure consistency of the effective description.\footnote{Some of the cosmological implications of this bound have been discussed in \cite{Scalisi:2018eaz,Cribiori:2021gbf,Scalisi:2024jhq,vandeHeisteeg:2023uxj, Lust:2023zql}. Stricter conditions arise when one applies instead the Higuchi bound, valid for spin 2 states of the tower, see $e.g.$~\cite{Noumi:2019ohm,Lust:2019lmq,Scalisi:2019gfv}.} A natural parametrization of the scalar potential is given by
\begin{equation}
\label{speciespotential}
V=M_p^{4-\alpha}(\Lambda_{sp})^\alpha\,, \qquad{\rm with}\qquad \alpha\geq 2\, .
\end{equation}
A scenario with such a potential satisfies the consistency requirement of having an infrared cut-off smaller than the species scale. 
Furthermore, this functional form \eqref{speciespotential} is supported by the anti-de Sitter distance conjecture~\cite{Lust:2019zwm}. 
By using the ansatz \eqref{speciesscalefinal}, we have then that the potential factorizes as
\begin{equation}
\label{eq:VgsR}
V(g_s, R) = \Lambda(R)\, V_q (g_s)\,,
\end{equation}
with dimensionful and dimensionless factors
\begin{equation}
\label{eq:LambdaVq}
\Lambda(R) = M_p^4 (M_p R)^{-\frac{\alpha p}{d+p-2}}\,, \qquad \text{and} \qquad V_q(g_s) = g_s^{\alpha q}=s^{-\frac{\alpha q}{2}}\,,
\end{equation}
where setting $s=1/g_s^{2}$ allows us to make direct contact with the previous section.\footnote{More in general, the quintessence field might also be one of the geometric moduli of a string compactification, for example a $T^2$ modulus, and the modular-invariant quintessence potential would then follow from the geometric duality symmetry of the compactification. Via string-string duality, the axio-dilaton field and such a geometric modulus could possibly be related to one another.} 
This represents a novel scenario for dark energy with a hybrid framework, where one field, $R$, sets the scale of the background energy density, while the field $s$ plays the role of quintessence and is responsible for its time variation.

We assume that the field $R$ is at the boundary of the moduli space, but we remain agnostic about the mechanism stabilizing it, if any. This assumption is sufficient to establish a connection with the Dark Dimension scenario, which links the smallness of today’s dark energy density to the presence of one \cite{Montero:2022prj} or two \cite{Anchordoqui:2025nmb} towers of dark Kaluza-Klein states (the case with two towers, however, implies a great amount of fine-tuning).  The time evolution of the dark energy density, as suggested by the latest DESI DR2 data, is then implemented through the dynamics of the field $s$ in the vicinity of the self-dual point, $s=1$, namely in the bulk. The field $s$ cannot lie in the vicinity of the boundary since no string tower has ever been observed at the neutrino scale. We have thus to deal with the challenge of promoting to the bulk an expression for the potential derived asymptotically. This step can be accomplished by exploiting modular symmetries, following the logic outlined in the previous section.

The radius $R$ of the large extra dimensions is setting the value $\Lambda \simeq 10^{-120} M_p^4$ of the dark energy density at the moment in which $s$ starts rolling. According to the Dark Dimension scenario, one has 
\begin{equation}
\Lambda(R) \simeq \frac{\lambda}{R^4}\,,
\end{equation} 
with $\lambda = \mathcal{O}(10^{-3})$ \cite{Anchordoqui:2022svl,Anchordoqui:2023oqm}. These assumptions set $R$ to be of order of the micron scale. From \eqref{eq:LambdaVq}, we can realize this scenario (namely, this $R$-dependence for $\Lambda$) when
\begin{equation}
d=4, \qquad \alpha = 4 \,\frac{p+2}{p}.
\end{equation}
In particular, for a single dark dimension one has $p=1$ and $\alpha = 12$, while for a double dark dimension, recently proposed in \cite{Anchordoqui:2025nmb}, $p=2 $ and $\alpha = 8$.

Concerning the quintessence field $s=1/g_s^2$, imposing that $V \to 1/s$ for $s \to \infty$ ($g_s \to 0$) sets $\alpha q =2$, due to equation \eqref{eq:LambdaVq}. Then, one has $q=1/6$ for $p=1$ and $q=1/4$ for $p=2$. The condition 
\eqref{relationgs} on the string coupling then reads
\begin{equation}
\label{relationgs1}
\frac{1}{M_p R} \leq g_s \leq \frac{1}{(M_p R)^\xi}\,,
\end{equation}
with $\xi = 2/5, 2/3$ in the first ($p=1$) and second ($p=2$) case respectively. 

This relation implies that at the boundary, for $g_s \to 0$, the complete potential $V(g_s,R)$, as given in \eqref{eq:VgsR}, falls off with a power of $R$ different from $1/R^4$. For example, assuming saturation of the lower bound in \eqref{relationgs1} one gets $V \simeq 1/R^6$, whereas assuming saturation of the upper bound gives $V \simeq 1/R^{\frac{24}{5}}$ for $p=1$ and $V \simeq 1/R^{\frac{16}{3}}$ for $p=2$. Instead, in the quintessence regime around the self-dual point, $g_s \simeq 1$, which does not lie in the region \eqref{relationgs}, one can have $V \simeq 1/R^4$.  
 
Let us now consider the strong coupling limit $g_s \to \infty$. We assume that also in this limit the ultraviolet cutoff of the effective description approaches zero. This means that there should exist a dual string scale, $\tilde M_s$, such that
\begin{equation}
\tilde M_s = \tilde g_s^{\frac{2}{d-2}} M_p = g_s^{-\frac{2}{d-2}}M_p\,.
\end{equation}
This scale is set by the tension of states becoming massless and that are non-perturbative with respect to $g_s$; candidates for such states are strings in four dimensions arising from wrapped (heterotic) NS5-branes \cite{Font:1990gx} with tension $T_5 \sim 1/g_s^2$. Adding the possible contribution of light Kaluza-Klein modes, we get a dual species scale
\begin{equation}
\tilde \Lambda_{sp} \simeq g_s^{-q} \, \frac{M_p}{(M_p R)^{\frac{p}{d+p-2}}}
\end{equation}
and potential
\begin{equation}
\tilde V(g_s, R) =V(1/g_s, R).
\end{equation}
This dual model, with the form of $\Lambda(R)$ dictated by the Dark Dimension scenario, is again realized by setting $d=4$, $\alpha q =2$, $\alpha p = 4(p+2)$, as above, and by changing $g_s\rightarrow 1/g_s$. 

Summarizing, we have considered a scenario where the dark energy density is described by a factorized potential, as given in \eqref{eq:VgsR} and \eqref{eq:LambdaVq}, and identified two viable possibilities depending on the number $p$ of (mesoscopic) extra dimensions that determine the present-day magnitude of such energy density. For $p=1$, one has $\alpha=12$ and $q=1/6$. For $p=2$, instead, $\alpha=8$ and $q=1/4$. This means that the case with a single extra dimension leads to a scalar potential characterized by a larger energy gap with respect to the species scale and faster decay to zero as the field $s$ increases, compared to the case with two extra dimensions. More importantly, we have pointed out that a framework with the magnitude of the dark energy density set by the compactification radius as $V\simeq 1/R^4$, as required by the Dark Dimension scenario, is only possible if $g_s\simeq 1$ and then the field $s$ lies in the bulk. In all other cases, in particular when $s$ approaches the boudaries ($s\rightarrow0$ or $s\rightarrow\infty$), the magnitude of the dark energy density is determined by different (negative) powers of $R$.

\subsection{Modular-invariant quintessence}

As anticipated, we would like to obtain an expression for the potential $V(g_s, R) = \Lambda(R) V_q(g_s)$ which we can trust also in the bulk of the moduli space, for what concerns the $g_s$ dependence. This is accomplished by implementing modular invariance on $V_q(g_s)$.  Proceeding step-wise, we can first glue the weak and strong $g_s$ coupling limit together. Specializing to $d=4$, $\alpha q =2$, $\alpha p = 4(p+2)$, we obtain self-dual expressions for the species scale
\begin{align}
\Lambda_{sp} &= \left(\frac{2}{s+1/s}\right)^{-q} \, \frac{M_p}{(M_p R)^{\frac{p}{p+2}}}
\end{align}
 and potential
\begin{align}
V(g_s, R) &= \left(\frac{2}{s+1/s}\right) \, \frac{1}{R^4},
\end{align}
which connects to the self-dual potential discussed in Sec.~\ref{Sec:ModPotential}.

Now, we can generalize to full modular invariance. We introduce the axion field $a$, describing the imaginary part of $S = s + ia$. Then, we employ a model for the S-modular invariant species scale discussed in  \cite{Cribiori:2023sch}. In this case, the number of species of string states as a function of $S$ is given by
\begin{equation}
N_{sp}(S,\bar S)\simeq -\log \lbrack|\eta(S)|^4(S+\bar S)\rbrack\, .
\end{equation}
One can check that for $s \to 0$ or $s\to\infty$, namely for large/small $g_s$,  one recovers $N_{sp} \simeq g_s^{-2}$. For the hybrid scenario, discussed above, $N_{sp}\simeq g_s^{-2q}$ as given by \eqref{eq:LambdaVq}. Then the complete general expression of modular invariant species scale can be  written as
\begin{eqnarray}
\Lambda_{sp}=\frac{M_p}{\left(-\log \lbrack|\eta(S)|^4(S+\bar S)\rbrack\right)^\frac{q}{2}    (M_pR)^{\frac{p}{p+2}}}\, ,
\end{eqnarray}
with $q=1/6$ for one dark dimension ($p=1$) and  $q=1/4$ for two dark dimensions ($p=2$).
The sought-after modular-invariant potential  $V(S,\bar S, R)$ is then
\begin{equation}\label{modpotential1}
V(S,\bar S,R)=-{1\over \log \lbrack|\eta(S)|^4(S+\bar S)\rbrack} \,\frac{1}{R^4}\, .
\end{equation}
One can check that for $s\rightarrow\infty$ this potential has the same scaling behavior as the self-dual potential in \eqref{Vinit}; in the saxionic direction $a=0$ it has also a very similar shape.

To summarize, via the potential \eqref{modpotential1} we embedded the educated guess \eqref{Vguess} into a more realistic scenario with two fields playing different roles: $R$ for the present value of the dark energy density and $S$ for its time evolution. This scenario is motivated by DESI DR2 data and also by general quantum gravity principles.

Concentrating only on the $S$-dependence a simple action functional invariant under modular transformations is
\begin{equation}
S_q =M_p^2 \int d^4 x\sqrt{-g}\left(- g_{S \bar S}\partial_\mu S \partial^\mu \bar S -M_p^2 V_q(S, \bar S)\right),
\end{equation}
with $g_{S \bar S} = n/(S+\bar S)^2$.
We can compute
\begin{equation}
M_p\sqrt{g^{S\bar{S}}\dfrac{\partial V_q}{\partial S}\dfrac{\partial V_q}{\partial \bar{S}}}\Bigg\vert_{\text{Re}S\to\infty}=\sqrt{2}/n,
\end{equation}
which matches the asymptotics of \eqref{Vinit} for $n=1$.  In fact,
any value $n>1$ would violate the TCC bound. Since any modular invariant function has vanishing derivative at the self dual point, we have also
\begin{equation}
 \sqrt{g^{S\bar{S}}\dfrac{\partial V_q}{\partial S}\dfrac{\partial V_q}{\partial \bar{S}}}\Bigg\vert_{S=1}=0\,,  \qquad M_p^2 \dfrac{   \text{ min}[\partial_{i}\partial_{j}V_q-\Gamma^k_{ij}\partial_k
     V_q]}{V_q}\Bigg\vert_{S=1}\simeq -7.2/n \,, 
\end{equation}
satisfying the refined de Sitter conjecture.

\vspace{0.5cm}

\section{Testing against DESI DR2 cosmological data}\label{Sec:DESI}

We now examine whether the modular-invariant potential \eqref{Vguess} is compatible with DESI constraints on late-time cosmic acceleration, particularly near the self-dual point. By combining BAO, CMB, and supernova data and modeling the dark energy equation of state as
$w(z) = w_0 + w_a \ z/(1 + z)$ with redshift $z$,
the DESI collaboration reported an unambiguous preference for a sector of the $(w_0,w_a)$ plane in which $w_0 + w_a < -1$
and \mbox{$w_0 > -1$}. This seems to suggest that $w(z)$
experiences a transition from a phase violating the null-energy condition at large $z$ to a phase obeying it at small
$z$~\cite{DESI:2025zgx}. However, this conclusion might be avoided since rather simple thawing quintessence models, satisfying the null energy condition for all $z$, predict an observational preference for the same sector~\cite{Shlivko:2024llw}. In this paper, we focus attention on thawing quintessence models that can accommodate the DESI DR2 BAO+CMB+SN results.

We compare the modular-invariant potential to the thawing quintessence model analyzed by the DESI collaboration with axion-like potential, 
\begin{equation}
  V (\phi) = m_a^2 \ f_a^2 \ [ 1 + \cos(\phi/f_a)] \,,
\label{Vaxion}
\end{equation}
where $m_a$ denotes the mass of the scalar and $f_a$ the decay constant, which is regarded as the effective energy~\cite{Freese:1990rb,Frieman:1995pm}. Using BAO data, CMB observations, and the three SN datasets, the DESI
collaboration~\cite{DESI:2025fii} reported the following constraints on the physical mass and the effective energy scale: 

\begin{subequations}
\label{DESIa}
\begin{align}
\log_{10}(m_a/{\rm eV}) &= -32.67^{+0.23}_{-0.25} && \qquad \text{BAO+CMB+PantheonPlus} \,, \\
\log_{10}(m_a/{\rm eV}) &= -32.50^{+0.28}_{-0.30} && \qquad \text{BAO+CMB+Union3} \,, \\
\log_{10}(m_a/{\rm eV}) &= -32.63^{+0.26}_{-0.30} && \qquad \text{BAO+CMB+DESY5} \,,
\end{align}
\end{subequations}
\vspace{-.7 cm}
\begin{subequations}
\label{DESIb}
\begin{align}
\log_{10}(f_a/M_p) &= -0.13^{+0.33}_{-0.29} && \qquad \text{BAO+CMB+PantheonPlus} \,, \\
\log_{10}(f_a/M_p) &= -0.29^{+0.63}_{-0.35} && \qquad \text{BAO+CMB+Union3} \,, \\
\log_{10}(f_a/M_p) &= -0.09^{+0.66}_{-0.40} && \qquad \text{BAO+CMB+DESY5} \,.
\end{align}
\end{subequations}\\
These constraints demand that the field starts in the hilltop regime, with initial condition 
\begin{equation}
\phi_i/f_a \sim 0.7 \, .
\label{phii}
\end{equation}
Then, the scalar field rolls down the potential, reaching the present value of $\phi_0/f_a \sim 1.1$, traversing approximately $\Delta \phi \sim 0.4 M_p$. 
Note that (\ref{phii}) is a result of a multi-parameter space likelihood analysis using the potential  (\ref{Vaxion}), with a prior on initial conditions of $\phi_i/f_a > 0.01$~\cite{Urena-Lopez:2025rad}.

 \begin{figure}
    \centering
        \begin{tabular}{cc}
          \includegraphics[width=0.48\linewidth]{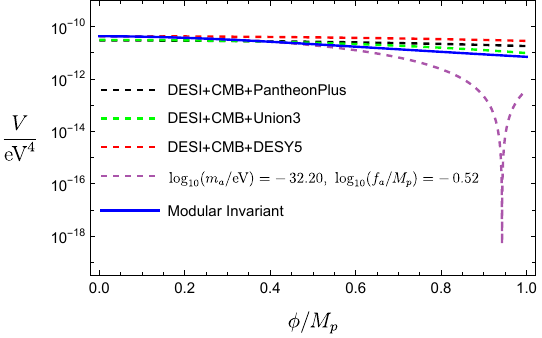}   & \includegraphics[width=0.49\linewidth]{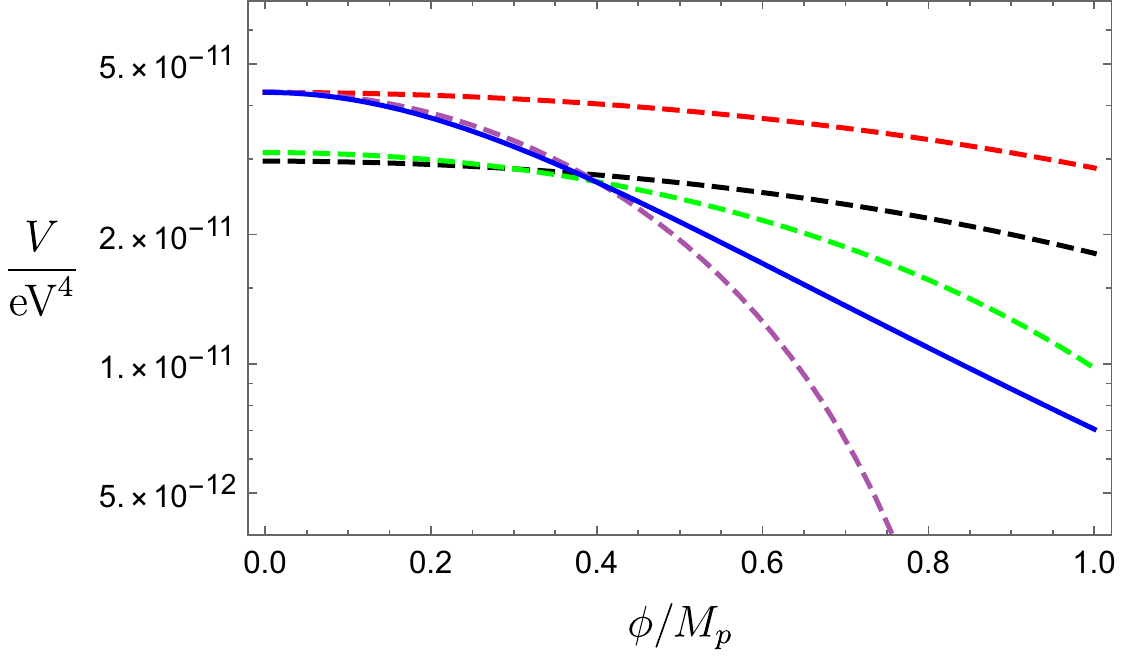}
        \end{tabular}

    \caption{Comparison of the modular-invariant potential \eqref{Vguess} along $a=0$ expressed in terms of the canonically normalized field $\phi = (1/\sqrt{2}) \log s$, with experimental results from \cite{DESI:2025zgx}.}
    \label{fig:Vquintessence}
\end{figure}

In figure~\ref{fig:Vquintessence}, we compare the
behavior of the modular invariant potential with the axion-like
potential \eqref{Vaxion} analyzed by the DESI collaboration at the central values of
the $1\sigma$ confidence intervals as given by \eqref{DESIa} and
\eqref{DESIb}. In the same figure, we also show a fit of the the axion-like potential to accommodate modular invariant potential. We can see that the best fit parameters,
\begin{equation}
  \log_{10}(m_a/{\rm eV}) = -32.20 \quad {\rm and} \quad \log_{10}(f_a/M_p) =
  -0.52 \,,
\label{logmi}  
\end{equation}
are consistent with DESI + CMB + Union3 at $1\sigma$, and are just outside the $1\sigma$ confidence regions of DESI + CMB + PantheonPlus and DESI + CMB + DESY5. 
An analogous analysis for the $S$-dual potential was performed in \cite{Anchordoqui:2025fgz}, where it was obtained $\log_{10}(m_a/{\rm eV}) = -32.47$ and  $\log_{10}(f_a/M_p) =-0.29$, whose values are also consistent at $1\sigma$  with DESI + CMB + Union3, DESI + CMB + PantheonPlus, and
 DESI + CMB + DESY5 \cite{Anchordoqui:2025fgz}.

Next, we investigate the behavior of the equation of state
parameter. We start from the S-dual potential, whose equation of state parameter was not analyzed in \cite{Anchordoqui:2025fgz}. Then, we look at the modular-invariant potential. To this end we study the dynamics of $\phi$ on the flat Friedmann-Lema\^{\i}tre-Robertson-Walker background with line element
\begin{equation}
ds^2 = -dt^2 + a^2(t) \left[dr^2 + r^2 d\Omega^2_2\right],
\end{equation}
where $a(t)$ is the scale factor with cosmic time $t$ and $r$ the comoving radial distance.
The equations of motion are (a dot represents a derivative with respect to $t$)
\begin{align}
&\ddot \phi + 3 H \dot \phi + V'(\phi) = 0 \,, 
\label{q1}\\
&3 \left(\frac{\dot a}{a}\right)^2  M_p^2 =  3 H^2 M_p^2 = 
\frac{1}{2} \dot \phi^2 + V(\phi) \equiv \rho_\phi,\\
&\frac{\ddot a}{a} = - \frac{1}{6 M_p^2} (\rho_\phi + 3 P_\phi) \, ,
\end{align}
while the dark energy equation of state is 
\begin{equation}
  w \equiv \frac{P_\phi}{\rho_\phi} = \frac{\dot \phi^2/2 -
    V(\phi)}{\dot \phi^2/2 + V(\phi)} \, .
\end{equation}  

\begin{figure}
    \centering
        \begin{tabular}{cc}
        \includegraphics[width=0.48\linewidth]{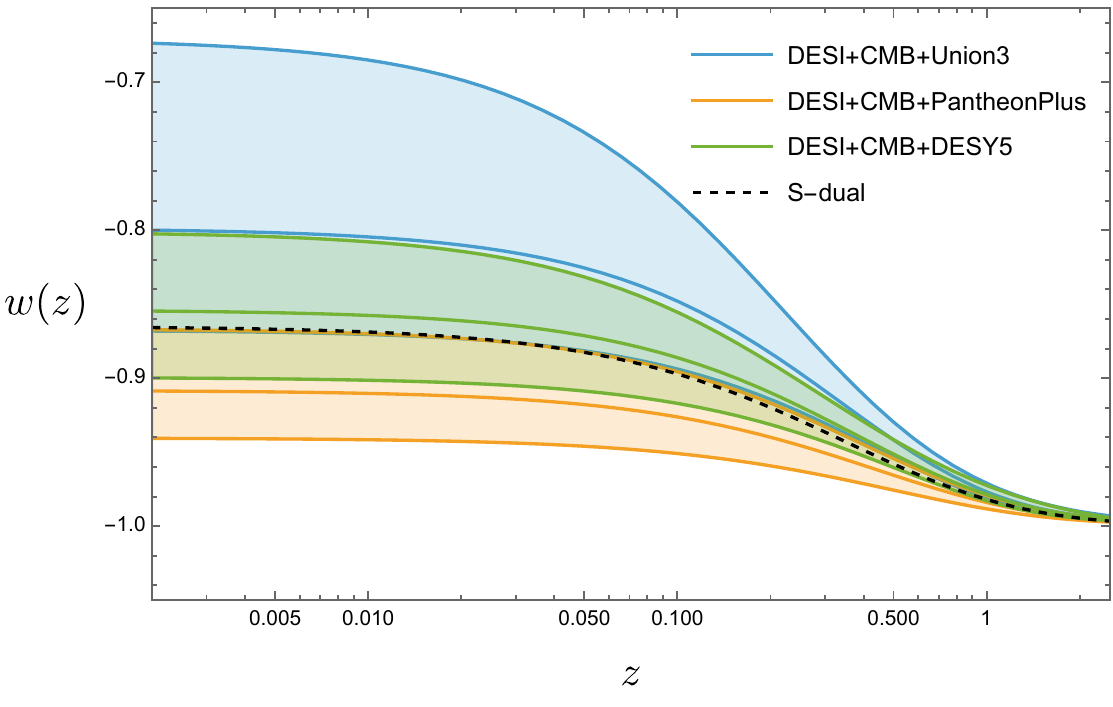}   & \includegraphics[width=0.48\linewidth]{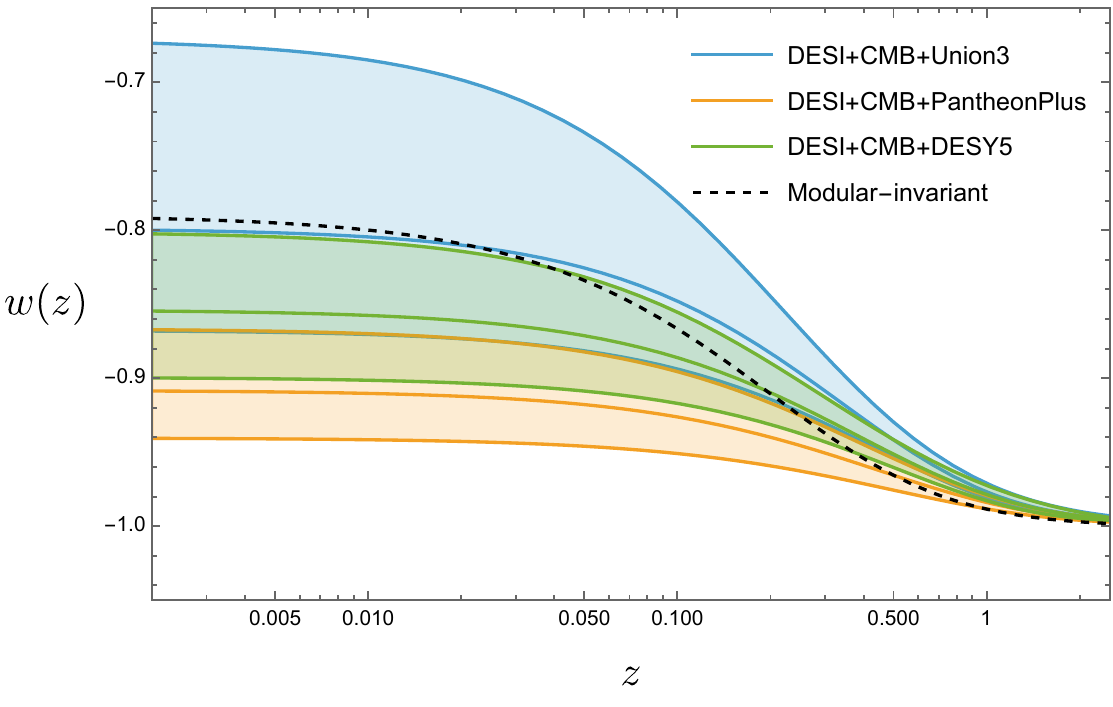}
        \end{tabular}
    \caption{The shaded regions show the marginalized constraints on
      the equation of state parameter, $w(z) = P_\phi/\rho_\phi$, as
      obtained by the DESI Collaboration for the axion-like potential
      given in eq.~(\ref{Vaxion})~\cite{DESI:2025fii}. The dashed lines show the analytic
      approximation to the equation of state parameter as given by
      eq.~(\ref{wanalytic}) for the $S$-dual potential (left) and the
      modular-invariant potential (right). We have taken $\Omega_\phi
      \sim 0.69$~\cite{DESI:2025zgx}, and we have set $w_0 = -0.87$ for
      the $S$-dual potential and $w_0 = -0.79$ for the modular invariant potential.}
    \label{fig:w}
\end{figure}

As for the S-dual potential, we expand it around the $\mathbb{Z}_2$-symmetric point $\phi = 0$ up to second order, 
$V(\phi) = \sum_{n=0}^2 V^{(n)} (0) \phi^n/n!$. For $|w + 1| \ll
1$, the evolution of the scale factor can be approximated by the one of $\Lambda$CDM cosmology
\begin{equation}
  a(t) = [(1 - \Omega_\phi)/\Omega_\phi]^{1/3} \sinh^{2/3}
  (t/t_\Lambda),
\label{aLCDM}
\end{equation}  
where $t_\Lambda = 2 M_p /\sqrt{3V(0)}$ and the density parameter
$\Omega_\phi$ is a measure of the ratio of the present-day dark energy density $\rho_{\phi_0}$ to the critical density $3 H_0^2 M_p^2$ required for the universe to be spatially flat, with $H_0$ is the Hubble constant. 
Substituting (\ref{aLCDM}) into (\ref{q1}) the equation of motion is integrated to give the field as a function of $t$ and so the field equation of state
\begin{equation}
  w \simeq -1 +\frac{\dot \phi}{\rho_\phi} \, .
\label{w1}
\end{equation} 
Then,  taking $\rho_\phi \sim V(0)$, we can rewrite \eqref{w1} as 
\begin{equation}
	w(z) = -1 + \frac{1+w_0}{(1+z)^{3(K-1)}}\tilde{F}^2(z),
\label{wanalytic}
\end{equation}
where 
\begin{align}
K & = \sqrt{1- \frac{4}{3} \ M_p^2 \
    \frac{V''(0)}{V(0)}} = \sqrt{\frac{11}{3}},\\
  \tilde{F}(z) &\equiv \frac{\left[K-F(z)\right]\left[1+F(z)\right]^K + \left[K+F(z)\right]\left[F(z)-1\right]^K}{(K-\Omega_{\phi}^{-1/2})(1+\Omega_{\phi}^{-1/2})^K+(K+\Omega_{\phi}^{-1/2})(\Omega_{\phi}^{-1/2}-1)^K}, \\
	F(z) &\equiv \sqrt{1+(\Omega_{\phi}^{-1}-1)(1+z)^3},
\end{align}
and where $w_0$ is a free parameter~\cite{Dutta:2008qn}. In the left panel of figure~\ref{fig:w} we compare the evolution of $w$ as given by \eqref{wanalytic} to the the marginalized posterior distribution for the equation of state parameter associated with the scalar field potential in \eqref{Vaxion}, obtained by the DESI Collaboration~\cite{DESI:2025fii} using BAO, CMB, and three SN compilations. In the right panel of figure~\ref{fig:w} we
do the same comparison but for the modular invariant potential, which has been approximated using (\ref{Vaxion}) with parameters given in \eqref{logmi}, and so $K = 2.9$.

In closing, we note that a recent study considering the same experimental data samples, but relaxing the prior on initial conditions leads to a steeper potential, with parameter $\log_{10}(f_a/M_p) = -0.5$ and
initial condition $\phi_i/M_p= 0.004$~\cite{Shlivko:2025fgv}. This study is more in line with the predictions of the modular-invariant potential. We conclude that the modular invariant potential, similarly to the $S$-dual scenario, provides a good representation of current observations.

\section{Comments on the supergravity embedding}\label{Sec:SUGRA}

In this section, we collect comments and attempts towards an embedding of the potential \eqref{modpotential1} in N=1 supergravity. We will explain the challenges to be overcome and finally we will present a supersymmetrization of \eqref{modpotential1} which exploits the versatility of non-linearly realized supersymmetry. Throughout the whole section we set $M_p=1$.

Modular-invariant supergravity models were constructed already in \cite{Ferrara:1989bc} and feature a constrained structure stemming from the interplay of supersymmetry and modular symmetry, see \cite{Cribiori:2024qsv} for a recent review. Crucially, their potential typically diverges asymptotically and thus cannot be identified with our proposal \eqref{modpotential1}.\footnote{A divergence is not acceptable for us because it would imply that the potential exceeds the species scale at weak string coupling, $s \to \infty$. In other words, the weak coupling limit would be obstructed.}
To simplify the problem, we aim first at obtaining S-duality invariant potentials. We will consider modular-invariant ones in a second step.

\subsection{Self-dual potentials}

To obtain a supergravity model in which the modular invariant potential vanishes asymptotically is not straightforward. Let us consider the general F-term potential 
\begin{equation}
\label{eq:VFG}
V_F = e^G (G^{i \bar \jmath}G_i G_{\bar \jmath} - 3), \qquad \text{with} \qquad G = K + \log W \bar W.
\end{equation}
A simple choice for the K\"ahler potential in the $S$-direction is 
\begin{equation}
\label{kp}
K(S, \bar S) = -\log (S+\bar S) = - \log (2s).
\end{equation}
One can check that a potential $V(s)\simeq 1/s$ for $s \to \infty$ is obtained for $W(S)= \sqrt 2$, while a dual potential $\tilde V(s) \simeq s$ for $s \to 0$ is obtained for $\tilde W(S) = \sqrt 2 S$. Even if we understand the asymptotics, to engineer a superpotential $W=W(S)$ reproducing the self-dual potential \eqref{Vinit} in the bulk it is not obvious. Indeed, since  $G$ must be invariant under $s\rightarrow 1/ s$, given the above choice of the K\"ahler potential it follows that superpotential must have duality weight -1, $i.e.$ it transforms as
\begin{equation}
W\to \frac{e^{i \varphi}W}{S},
\end{equation}
where $\varphi$ is an arbitrary multiplier phase.

One superpotential with this property is $W(S) = \sqrt{\frac{2S}{S+1/S}}$. Assuming that a separate no-scale sector cancels the negative term in \eqref{eq:VFG} (related to the gravitino mass), the resulting potential is $V(s)=s(s^2-1)^3/(s^2+1)^3$. This falls as $1/s$ for $s \to \infty$, but has a minimum at $s=1$ and two maxima at $s=\sqrt{5\pm2 \sqrt{6}}$, so it does not match with \eqref{Vinit}. 
Another superpotential with the correct modular behavior is $W=i+S$. This can arise as a type IIB  flux superpotential $W(S)=F_3+iSH_3$, where the NS 3-form flux $H_3$ gets exchanged with the RR 3-form flux $F_3$ under duality transformations. Assuming again that the gravitino term is taken care of by a separate no-scale sector we have $V(s) = (s^2+1)/(2s)$. This potential grows linearly at large $s$ but has a single extremum (minimum) at $s=1$; in fact, it is the inverse of \eqref{Vinit}. 

Let us now discuss two examples with a more positive outcome. In the first example, we consider the superpotential
\begin{equation}
W(S) = \sqrt{\frac{S}{2}}\left( \arctan{S} - {\rm arccot}\,{S}\right).
\end{equation}
Together with \eqref{kp}, this results in an S-duality invariant model.\footnote{A closely related superpotential leading to the same scalar potential is $W(S)=\sqrt{\frac S2}\log\left(\frac{1+i S}{1-iS}\right)$.} The associated scalar potential, assuming once more a no-scale cancellation of the gravitino term due to a separate sector, becomes (setting $a=0$)
\begin{equation}
V(s) = \frac{4s^2}{(1+s^2)^2}\, .
\end{equation}
This is the square of the self-dual potential \eqref{Vinit}. Connecting to the discussion in the previous section, this potential is approximated by an axion-type potential with parameters $\log_{10}(m_a/eV) = -32.25$ and $\log_{10} (f_a/M_p) = -0.51$. The initial condition is $\phi_i/M_p = 0.22$ and the field today is at  $\phi_0/M_p = 0.34$. The approximation of the axion-like potential is relatively good for these values, so this supergravity model is consistent with DESI data.

In the second example, we follow the prescription of \cite{Scalisi:2015qga} (see \cite{DallAgata:2014qsj,Kallosh:2014hxa} for similar constructions in flat K\"ahler geometry) and make the following ansatz for the superpotential
\begin{equation}
    W=\sqrt{\frac{S}{2}}\int \dfrac{\sqrt{\mathcal{V}(S)}}{S}dS\,, \qquad \mathcal{V}(S) = \frac{2}{S+1/S}\,,
\end{equation}
where the integration constant must be fixed to have $W$ constant at infinity. After the usual no-scale cancellation of the gravitino term, one can check that the potential of the model, for $a=0$, reproduces precisely the self-dual potential \eqref{Vinit}.

An alternative method to ensure a positive definite potential $V_F$ and to avoid runaway directions, usually associated with no-scale symmetries in multi-field setups, is to use a nilpotent superfield $X$ (as it will be introduced in the next sub-section) with a canonical K\"ahler potential and a linear superpotential, see $e.g.$ \cite{DallAgata:2014qsj,Kallosh:2014hxa,Scalisi:2015qga}.

\subsection{Modular-invariant  potential}

When passing from S-duality to full modular-invariance, the superpotential must transform as a modular function of weight $-1$ \cite{Ferrara:1989bc,Font:1990nt,Font:1990gx}, $i.e.$
\begin{equation}
\label{WtransfSL2Z}
W(S)\rightarrow {e^{i\varphi}\over icS+d} W\, .
\end{equation}
If we insist on the simple choice of K\"ahler potential \eqref{kp}, we can argue that recovering \eqref{Vguess} is not possible. We can proceed by contradiction. Using \eqref{kp}, we can rewrite \eqref{Vguess} as $V^{-1}= K-Z(S)-\bar Z(\bar S)= G-\log W-\log \bar W -Z(S)-\bar Z(\bar S)$, with $Z(S) = 2 \log \eta(S)$. Now, we use that $V$ must be invariant under K\"ahler transformations. $G$ is invariant, while $W$ transforms as $W\to e^{-F} W$, with $F$ an arbitrary holomorphic function. To cancel the K\"ahler variation of $W$, $Z$ must transform as well. However, since $F$ is arbitrary while $Z$ is given, the only option is to impose $Z=-\log W$, or equivalently $W = 1/\eta(S)^2$. One can then check that the scalar potential, associated to \eqref{kp} and to this choice of $W$, is not \eqref{Vguess}. For example, this potential diverges asymptotically. Hence, to recover \eqref{Vguess} it is necessary (but it might not be sufficient) to abandon the assumption that \eqref{kp} is the appropriate K\"ahler potential. 

Given the difficulty associated to the F-term potential, one might try with a D-term. Again, we can argue that the simple choice of K\"ahler potential \eqref{kp} would not allow us to recover \eqref{Vguess}. D-terms are associated to local symmetries of the scalar manifold. We can gauge only those symmetries which appear as global isometries and the latter are fixed by the choice of the metric. In particular, given the metric stemming from the K\"ahler potential \eqref{kp}, we can only gauge the shift symmetry $S \to S+ i\theta$, $\theta \in \mathbb{R}$. The associated holomorphic killing vector is constant, $k=i$, and its moment map is thus such that $\partial_{\bar S}P = -g_{S \bar S}$, $i.e.$ $P=-K_{\bar{S}}-i \,r(S)$. Since $K$ is gauge-invariant, $r=i \xi$, with $\xi \in \mathbb{R}$ a Fayet-Iliopoulous term. To write the D-term potential, $V_D$, we need to specify a gauge kinetic function $f=f(S)$ such that  $V_D = \frac12 ({\rm Re} \,f)^{-1} P^2$. 
Concretely, we need ${\rm Re }\, f \simeq P^2/V$ with $P^2 = (1/(S+\bar S)+\xi)^2$ and $V$ given by \eqref{Vguess}. However, one can check that this function is not harmonic and thus it cannot be the real part of an holomorphic $f$. Hence,  the potential \eqref{Vguess} cannot be reproduced via a D-term in this way. Once more, it is necessary (but it might not be sufficient) to abandon the assumption that \eqref{kp} is the appropriate K\"ahler potential.

A possible way out to the above difficulties is to consider a regime in which supersymmetry is non-linearly realized. This might be justified since the potential \eqref{Vguess} breaks supersymmetry at any point and there is no phase in which it is restored. In this sense, it is natural to describe \eqref{Vguess} in a theory with spontaneously broken and non-linearly realized supersymmetry.  
While non-linear supersymmetry allows for the supersymmetrization of any desired bosonic model, an efficient approach involves constrained multiplets \cite{Rocek:1978nb,Casalbuoni:1988xh,DallAgata:2016syy,Cribiori:2017ngp} (see \cite{Antoniadis:2024hvw} for a recent review), thanks to which one can have some hints on the model in the linearly-realized regime. In this approach, without loss of generality the non-linear realization can be implemented via a chiral multiplet $X$ that, in order to propagate only the goldstino degrees of freedom, is assumed to be nilpotent, $X^2=0$. With this basic ingredient, one can employ the standard language of linearly realized supergravity to describe the desired model. In our case, following \cite{Antoniadis:2014oya,McDonough:2016der,Kallosh:2017wnt,Farakos:2018sgq}, we propose that the scalar potential \eqref{Vguess} is described by the K\"ahler-invariant function
\begin{equation}
G = -\log g -\frac{1}{g\,V},\qquad \text{with} \qquad g(S, \bar S, X, \bar X) =\frac{S+\bar S}{X \bar X}
\end{equation}
and where $V=V(S, \bar S)$ is the desired modular-invariant potential \eqref{Vguess}. An equivalent form in terms of $K$ and $W$ is
\begin{equation}
    K = -\log (S+\bar S)-\frac{X \bar X}{(S+\bar S) V(S, \bar S)}, \qquad W= X.
\end{equation}
By direct computation, one can check that at $X=0$ this model gives an F-term potential $V_F = V(S, \bar S)$. Since $V$ is invariant by assumption, modular invariance of the full model is achieved if the function $g$ is modular invariant. This is so if $X$ transforms as 
\begin{equation}
X \to \frac{ e^{i \varphi}X}{ic S+d},
\end{equation}
in accordance with \eqref{WtransfSL2Z}. 
Hence, non-linear supersymmetry allows for a supersymmetrization of the modular-invariant potential while at the same time preserving modular invariance. To explain the microscopic origin of this model remains as an open problem.

\section{Discussion}\label{Section 6}

In this article, motivated by the latest DESI DR2 results and by general principles of quantum gravity, we  proposed and studied a novel scenario, where the dark energy density evolves dynamically. We obtained two main results:
\begin{itemize}

\item First, we proposed a modular-invariant quintessence potential that satisfies the swampland constraints and is consistent with the recent observations by DESI. This modular-invariant potential is inspired, and in fact bounded, by a modular-invariant function arising as the species scale in certain string compactifications. 
\item Second, we introduced a new framework for dark energy, in which one scalar field is close to the boundary of  moduli space, while a second field resides in the bulk.\footnote{In \cite{Lust:2024aeg}    an index was proposed that relates potential on its boundary to the critical points in
its interior.}
The first field is a compactification radius $R$ of the extra-dimensions and, via a potential $\Lambda\simeq R^{-4}$, is responsible for the current magnitude of the dark energy density. The second field $s$ is related to the string coupling $g_s$ and, via the modular-invariant potential as given by \eqref{Vintro}, determines the quintessence time-variation of dark energy.\footnote{The time variation of the string coupling constant may lead to a time variation of standard model gauge coupling constants and potentially violates 
some strong bounds on the time variations of these couplings. However one can show that in type IIB compactifications with the standard model gauge groups localized on D7-branes or in type IIA 
compactifications with the standard model gauge groups localized on intersecting D6-branes, the gauge couplings are not given by the S-field however by the geometric moduli $T$ or $U$ respectively and therefore these bounds can be evaded.}

\end{itemize}

\noindent The proposal has a number of attractive features. First, the quintessence phase occurs close to the self-dual point ($s\simeq1$), where $g_s$ is of order one. This is important as, at present, the string coupling $g_s$ cannot be small, otherwise light string states would have been observed. Second, it predicts that, in the future, as the quintessence field rolls down to weak coupling, exponentially light string states should appear. Third, the decrease of the energy density, during the quintessence phase, is accompanied by a fall-off of the species scale $\Lambda_{sp}$, albeit only by a small amount. This is in agreement with the laws of species thermodynamics proposed in \cite{Cribiori:2023ffn}. Finally, our hybrid framework offers further support for a key feature of the Dark Dimension scenario: the observation that the inverse fourth power of $R$ (with $R$ on the order of a micron) naturally sets the present-day value of the dark energy density. In sec.~\ref{Sec:HybScenario}, we indeed showed that a potential such as $V\simeq R^{-4}$ is only possible if $g_s=\mathcal{O}(1)$, that is, the field $s$ lies close to the bulk. In other situations, for other magnitudes of the string coupling, the dark energy density $V$ will be related to other (negative) powers of $R$.

There are multiple directions in which this work can be extended in the future. While we mainly concentrated on a single modular-invariant potential, much more can be constructed. It would thus be interesting to test them against experimental observations. Besides, it is of crucial importance to obtain a deeper understanding, from a microscopic point of view, for any of these potentials, as well as for our bulk/boundary scenario.

It remains to be seen if the experimental results from DESI DR2 will become statistically more significant, or if the $\Lambda$CDM concordance model will eventually prevail.
Nevertheless, from a quantum gravity perspective we believe that a time-dependent dark energy, as in the scenario here proposed, is more favored over a (meta-)stable de Sitter universe.

\section*{Acknowledgements}

We are thankful to I.~Basile, A.~Bedroya, F.~Farakos and C.~Vafa for valuable discussions.
The work of L.A.A. is supported by the U.S. National Science
Foundation (NSF Grant PHY-2412679). I.A. is supported by the Second
Century Fund (C2F), Chulalongkorn University. 
The work of N.C. is supported by the Research Foundation Flanders (FWO grant 1259125N). 
The work of D.L. is supported by the Origins Excellence Cluster and by the German-Israel-Project (DIP) on Holography and the Swampland. J.M. thanks the Department of Physics and Astronomy ``Ettore Majorana'' at the University of Catania for kind hospitality during the completion of part of this work.

\bibliographystyle{JHEP}
\bibliography{refs}  

\end{document}